\documentclass[%
 reprint,
 amsmath,amssymb,
 aps,
]{revtex4-2}

\usepackage{xcolor}
\usepackage{graphicx}% Include figure files
\graphicspath{{./Figs/}}
\usepackage{dcolumn}% Align table columns on decimal point
\usepackage{bm}% bold math
\begin{document}
\preprint{APS/123-QED}

\title{Dam break of viscoplastic elliptical objects}% Force line breaks with \\

\author{Kindness Isukwem}
\thanks{kindness-chinwendu.isukwem@minesparis.psl.eu}
\affiliation{Mines Paris, PSL University, Centre for material forming (CEMEF), UMR CNRS 7635, rue  Claude Daunesse, 06904 Sophia-Antipolis, France}

\author{Anselmo Pereira}
\thanks{anselmo.soeiro\_pereira@minesparis.psl.eu}
\affiliation{Mines Paris, PSL University, Centre for material forming (CEMEF), UMR CNRS 7635, rue  Claude Daunesse, 06904 Sophia-Antipolis, France}%

\date{\today}% It is always \today, today,

%%%%%%%%%%%%%%%%%%%%%%%%%%%%%%%%%%%%%%%%%%%%%%%%%%%%%%%%%%%%%%%%%%%%%%%%%%%%%%
%%%%%%%%%%%%%%%%%%%%%%%%%%%%%%%%%%%%%%%%%%%%%%%%%%%%%%%%%%%%%%%%%%%%%%%%%%%%%%
\begin{abstract}
In this note, we numerically and theoretically analyze the physical mechanisms controlling the gravity-induced spreading of viscoplastic elliptical metric objects on a sticky solid surface (without sliding). The two-dimensional collapsing objects are described as Bingham fluids. The numerical simulations are based on a variational multi-scale approach devoted to multiphase non-Newtonian fluid flows. The results are depicted by considering the spreading dynamics, energy budgets, and new scaling laws. They show that, under negligible inertial effects, the driving gravitational energy of the elliptical columns is dissipated through viscoplastic effects during the collapse, giving rise to three flow regimes: gravito-viscous, gravito-plastic, and mixed gravito-visco-plastic. These regimes are strongly affected by the initial aspect ratio of the collapsing column, which reveals the possibility of using morphology to control spreading. Finally, the results are summarized in a diagram linking the object's maximum spreading and the collapse time with different collapsing regimes through a single dimensionless parameter called \textit{collapse number}.

\vspace*{0.9cm}

 
\end{abstract}
%%%%%%%%%%%%%%%%%%%%%%%%%%%%%%%%%%%%%%%%%%%%%%%%%%%%%%%%%%%%%%%%%%%%%%%%%%%%%%
%%%%%%%%%%%%%%%%%%%%%%%%%%%%%%%%%%%%%%%%%%%%%%%%%%%%%%%%%%%%%%%%%%%%%%%%%%%%%%

\maketitle

%%%%%%%%%%%%%%%%%%%%%%%%%%%%%%%%%%%%%%%%%%%%%%%%%%%%%%%%%%%%%%%%%%%%%%%%%%%%%%
%%%%%%%%%%%%%%%%%%%%%%%%%%%%%%%%%%%%%%%%%%%%%%%%%%%%%%%%%%%%%%%%%%%%%%%%%%%%%%
%--------------------------------------------------------------------------------------------------------------------------------------------------------------------------------------%--------------------------------------------------------------------------------------------------------------------------------------------------------------------------------------
\section{Introduction} \label{INTRO}
%--------------------------------------------------------------------------------------------------------------------------------------------------------------------------------------%--------------------------------------------------------------------------------------------------------------------------------------------------------------------------------------
The gravitational collapse of a complex fluid column is a paramount topic in Fluid Mechanics \citep{Liu-18, Valette-21}. It is related to a wide range of natural and applicative issues, from geophysical to industrial contexts, which includes preventing disasters caused by mud surges and the sudden collapse of mine tailing deposits, and controlling the release of fluids in non-conventional rheometric tests, such as the slump test for concrete and the Bostwick test for food \citep{Pashias-96, Schowalter-98, Roussel-05, Balmforth-07}. In such flow scenarios, many complex fluids behave as ideal (non-deformable) solids at low-stress levels and non-Newtonian liquids at stress levels above their yield stress $\tau_0$, thus termed viscoplastic fluids \citep{Balmforth-14}. Definitively, micro-structural interactions developed in these fluids (such as intermolecular attractive/repulsive forces, friction, capillary bridge, etc.) are translated macroscopically through a yield stress and a strain-rate-dependent viscosity $\eta$. This viscosity is often represented by the Bingham equation as $\eta= k+ \tau_0/|\boldsymbol{\dot{\gamma}}|$, where $k$ is the consistency index and $|\boldsymbol{\dot{\gamma}}|$ is the norm of the strain-rate tensor \citep{Bingham-16, Bingham-22}. Most fluids have been seen to behave like viscoplastic fluids under certain flow conditions. They include mineral suspensions (bentonite, kaolin, carbon black, etc.), organic suspensions/gels (Carbopol, alginate, ketchup, etc.), emulsions (mayonnaise) and foams \citep{Coussot-07, Cohen-Addad-13, Isukwem-24a}. 

Gravity-induced flows of viscoplastic fluid columns can be substantially affected by $\tau_0$ \citep{Liu-16, Valette-21, Liu-18}. When sufficiently high, the yield stress can hold the fluid up against gravity, either preventing the onset of the collapse or stopping the fluid flow due to the development of localised rigid-like regions commonly named \textit{plugs} \citep{Valette-21}. Interestingly, the growth of such regions is highly affected by morphological aspects of the fluid column, a topic scarcely explored in the literature, which has mainly focused on cylinders and prisms \citep{Liu-18, Valette-21}. Hence, it is reasonable to extend these studies to include other morphologies, for example, objects that can be considered two-dimensional (objects whose width $\gg$ height and thickness, for instance). In that sense, one key question is: can we predict the maximum spreading reached by a collapsing viscoplastic 2D column under gravity effects and the spreading time? We provide an answer to this question in the present work.  

In this short communication, we present a theoretical and numerical study devoted to the physical mechanisms driving the gravity-induced collapse of viscoplastic elliptical objects on a sticky solid surface. More specifically, we extend previous cylindrical and prismatic-based works \citep{Liu-18, Valette-21} by analyzing the spreading dynamics, maximum spreading, and spreading time of viscoplastic elliptical columns (2D). We employ 2D numerical simulations based on a variational multi-scale approach devoted to non-Newtonian multiphase flows. The results are analyzed by considering the spreading dynamics, energy budgets, and new scaling laws. Lastly, they are summarized in a two-dimensional diagram linking the maximum spreading and its spreading time with different flow regimes through a single dimensionless parameter called here the \textit{collapse number}. 

The organisation of the paper is as follows. A detailed description of the physical formulation and the numerical method is presented in section \ref{PFNMDN}. Numerical results are discussed in section \ref{RD}. Finally, conclusions and perspectives are drawn in the closing section.   

%%%%%%%%%%%%%%%%%%%%%%%%%%%%%%%%%%%%%%%%%%%%%%%%%%%%%%%%%%%%%%%%%%%%%%%%%%%%%%
%%%%%%%%%%%%%%%%%%%%%%%%%%%%%%%%%%%%%%%%%%%%%%%%%%%%%%%%%%%%%%%%%%%%%%%%%%%%%%
\section{Physical Formulation, Numerical Method and Dimensionless Numbers} \label{PFNMDN}
%%%%%%%%%%%%%%%%%%%%%%%%%%%%%%%%%%%%%%%%%%%%%%%%%%%%%%%%%%%%%%%%%%%%%%%%%%%%%%
%%%%%%%%%%%%%%%%%%%%%%%%%%%%%%%%%%%%%%%%%%%%%%%%%%%%%%%%%%%%%%%%%%%%%%%%%%%%%%

\begin{figure*}%[h!]
\centering
\includegraphics[angle=0, scale=0.20]{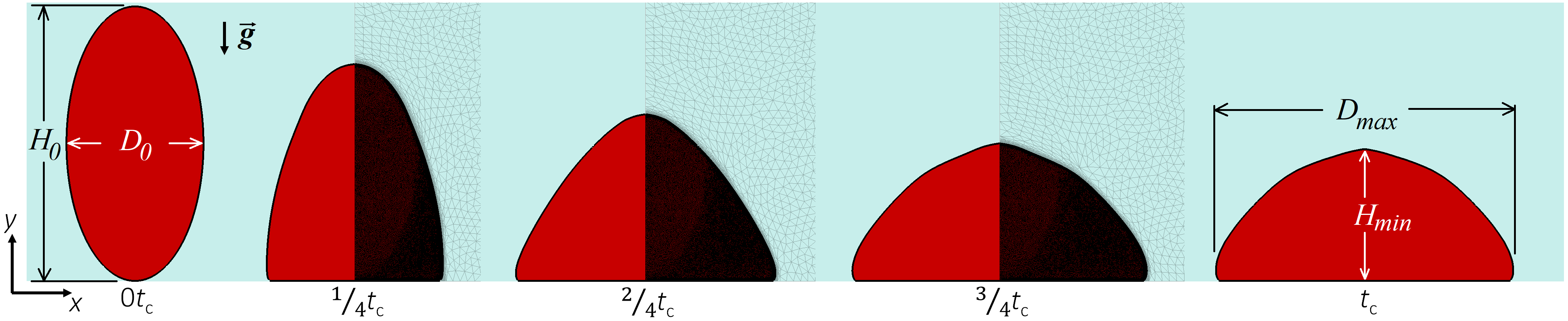}
\vspace*{-0.5cm}
\caption{Multiphase numerical simulation: a typical collapse and spreading of a viscoplastic elliptical object (shown in red) of initial height $H_0$, initial thickness $D_0$, density $\rho$, consistency $k$, and yield stress $\tau_0$ on a sticky surface. After the onset of the collapse at time $t_0$, the object spreads until it reaches its maximum spreading $D_{max}$ and a minimum height $H_{min}$ at time $t_c$. Both the 2D object and the solid surface are surrounded by air (depicted in blue). The mesh used in the simulation, shown with black and gray lines, consists of approximately 10$^5$ elements, with a minimum size of 1$\mu$m. A movie showing a typical numerical simulation is available in which $H_0 = 1$m, $D_0 = 0.5$m, $\rho =$ 0.001kg/m$^3$, $k =$ 1Pa$\cdot$s, $\tau_0 =$ 0.38Pa, $|\boldsymbol{g}| =$ 1000m/s$^2$ (the colors indicate the norm of the velocity vector $|\boldsymbol{u}|$). ({https://youtu.be/aCiPFHAmWBE})}
\label{fig-1}
\end{figure*}

As previously mentioned, we study the spreading of viscoplastic elliptical objects with metric dimensions on a no-slip solid surface, as depicted in figure \ref{fig-1}. These objects, characterized by an initial height $H_0$, thickness $D_0$, density $\rho$, consistency $k$, and yield stress $\tau_0$, collapse under the influence of gravity $\boldsymbol{g}$. Following the collapse, they spread slowly across the solid surface as their gravitational potential energy is dissipated through viscoplastic effects. The collapse continues until the objects reach a maximum spreading $D_{max}$ and a minimum height $H_{min}$ at time $t_c$. Figure \ref{fig-1} illustrates the spreading at four different times, ranging from $t/t_c=0$ (onset of collapse) to $t/t_c = 1$ (stoppage). The 2D objects and the solid surface are surrounded by air, which has a density $\rho_{air}$ and viscosity $\eta_{air}$. 

Our computational method is based on a massively parallel finite element library (CIMLIB-CFD) devoted to non-Newtonian multiphase flows \citep{Valette-19, Pereira-19, Pereira-20, Valette-21, Isukwem-24a, Isukwem-24b, Isukwem-24d, Isukwem-24e}. More specifically, we apply the momentum conservation equation presented below to the solenoidal flow ($\boldsymbol{\nabla \cdot u} = 0$) described earlier (figure \ref{fig-1}):
\begin{equation}
\rho \left( \frac{\partial \boldsymbol{u}}{\partial t} + \boldsymbol{u} \cdot \nabla \boldsymbol{u} - \boldsymbol{g} \right) = - \nabla p + \nabla \cdot \boldsymbol{\tau} \, ,
\label{eq:momentum}
\end{equation}
where $\boldsymbol{u}$ is the velocity vector, $\nabla$ denotes the gradient operator, $\boldsymbol{g}$  represents the gravitational acceleration vector, $p$ is the pressure, and $\boldsymbol{\tau}$ is the extra-stress tensor. The extra-stress tensor is defined as $\boldsymbol{\tau} = \eta \boldsymbol{\dot{\gamma}}$, where $\boldsymbol{\dot{\gamma}}$ represents the rate-of-strain tensor, given by $\boldsymbol{\dot{\gamma}} = \left( \boldsymbol{\nabla u} + \boldsymbol{\nabla u}^T \right)$. Our 2D objects are modelled as Bingham fluids, with viscosity $\eta$ described by: 
\begin{equation}
\eta = k + \frac{\tau_0}{{\mid \boldsymbol{{\dot{\gamma}}} \mid}}\left( {1-e^{-{\mid \boldsymbol{{\dot{\gamma}}} \mid}/{\dot{\gamma_p}}}}  \right)  \, .
\label{eq:bingham}
\end{equation}

As indicated in equation \ref{eq:bingham}, the viscosity incorporates the Papanastasiou regularization term, shown in parentheses, which prevents numerical divergence when the norm of the rate-of-strain tensor, ${| \boldsymbol{{\dot{\gamma}}}|}$, approaches zero \cite{Papanastasiou-87}. This regularization ensures that the viscosity plateaus when ${|\boldsymbol{{\dot{\gamma}}}|}$ falls below ${\dot{\gamma_p}}$, where ${\dot{\gamma_p}} = 10^{-6}$s$^{-1}$.

The numerical methods used here are based on a \textit{Variational Multiscale Method} (VMS) coupled with an anisotropic mesh adaptation method \citep{Valette-19, Pereira-19, Pereira-20, Valette-21, Isukwem-24a, Isukwem-24b, Isukwem-24d, Isukwem-24e}. The meshes used are composed of approximately 10$^5$ elements whose minimum size is 1$\mu$m (see the black lines in figure \ref{fig-1}). The time-dependent evolution of the 2D object's interface is captured using a Level-Set function. It is important to emphasise that our numerical multi-phase framework has been validated for a variety of flow scenarios, which includes the impact of drops \citep{Isukwem-24a, Isukwem-24b, Isukwem-24d, Isukwem-24e}, dam breaks \citep{Valette-21}, the stretching and breakup of filaments \citep{Valette-19}, and the development of buckling instabilities \citep{Pereira-19, Pereira-20} using Newtonian and non-Newtonian fluids. 

The 2D collapsing objects are on a metric scale of $H_0 = 1$m, and the aspect ratios of $H_0/D_0$ range from 0.25 to 10. Gravity $|\boldsymbol{g}|=$ 1000m/s$^2$. A yield-stress spectrum ranging from 10$^{-5}$Pa to 0.4Pa is considered, while the consistency and the density of the elliptical columns are held constant at $k=1$Pa$\cdot$s and $\rho=$10$^{-3}$kg/m$^3$, respectively. The viscosity and density of the surrounding air are set at $\eta_{air}=10^{-5}$Pa$\cdot$s and $\rho_{air}=$1kg/m$^3$. Additionally, a no-slip condition is imposed between the collapsing objects and the solid surface, with zero normal stresses on the walls of the computational domain \citep[similar to][]{Valette-21}. A movie showing a typical numerical simulation is available in which $H_0 =$ 1m, $D_0 =$ 0.5m, $\rho =$ 0.001kg/m$^3$, $k =$ 1Pa$\cdot$s, $\tau_0 =$ 0.38Pa, $|\boldsymbol{g}| =$ 1000m/s$^2$ (the colors indicate the norm of the velocity vector $|\boldsymbol{u}|$). ({https://youtu.be/aCiPFHAmWBE})

The dimensionless numbers that govern the problem denoted as $\Pi_i$, are derived from the Buckingham-$\Pi$ theorem. The relevant variables considered are $H_0$, $D_0$, $\rho$, $k$, $\tau_0$, $\boldsymbol{g}$ and the characteristic collapse velocity $U_c$ (which will be deduced in the following lines). The fundamental units involved are mass [kg], length [m], and time [s]. This analysis leads to four significant dimensionless quantities:  
\begin{equation}
\Pi_1 = \frac{H_0}{D_0}  \, ,
\label{eq:pi-1}
\end{equation}
\begin{equation}
\Pi_2 = \frac{k \left(U_c/D_0 \right)}{\rho g H_0}  \, ,
\label{eq:pi-2}
\end{equation}
\begin{equation}
\Pi_3 = \frac{\tau_0}{\rho g H_0}  \, ,
\label{eq:pi-3}
\end{equation}
\begin{equation}
\Pi_4 = \frac{\rho U_c^2}{\rho g H_0}  \, ,
\label{eq:pi-4}
\end{equation}
where $\Pi_2 = 1/\mathrm{Ga_k}$ ($\mathrm{Ga_k}$ is the $k$-based Galileo number), $\Pi_3 = \mathrm{Pl}$ ($\mathrm{Pl}$ is the plastic number) and $\Pi_4 = \mathrm{Fr}$ ($\mathrm{Fr}$ is the Froude number). Given that the flows considered here are inertia-free (this will be shown in detail in Section \ref{RD} through energy budgets), $\rho U_c^2$ is negligible when compared to the gravitational stress $\rho g H_0$ (typically $\Pi_4 \lesssim 10^{-5}$). Consequently, our flow scenarios can be described with the aid of three dimensionless numbers:  
\begin{equation}
\frac{H_0}{D_0}  ~~~ \text{(aspect ratio)}\, ,
\label{eq:aspect-ratio}
\end{equation}
\begin{equation}
\mathrm{1/Ga_k} = \frac{k \left(U_c/D_0 \right)}{\rho g H_0}  ~~~ \text{(inverse Galileo number)}\, ,
\label{eq:inv-Ga}
\end{equation}
\begin{equation}
\mathrm{Pl} = \frac{\tau_0}{\rho g H_0}  ~~~ \text{(plastic number)}\, .
\label{eq:Pl}
\end{equation}
The impact of these three dimensionless numbers on the collapse of viscoplastic 2D columns is examined in section \ref{RD}. Specifically, the results are framed and analyzed in terms of these dimensionless parameters, providing a clear understanding of how each one influences the spreading dynamics.

Finally, a characteristic collapse velocity, $U_c$, can be defined by considering that the gravitational energy (per unit length) of the elliptical fluid object at $t = 0$s ($\rho g H_0 \frac{\pi H_0 D_0}{4}$) is converted into kinetic energy [$\frac{\rho U_c^2}{2} \frac{\pi H_0 D_0}{4}$], and dissipated [$4 k \frac{U_c}{H_0} H_0 D_0 + 2 \tau_0 H_0 D_0$, from the column compression standpoint] throughout the collapse process. The factors 4 and 2 in the viscous and plastic dissipation terms account for the planar compression along the 2D column, which will be further illuminated in section \ref{RD}. This yields the following equation:
\begin{equation}
\frac{\pi}{4} \frac{\rho}{2} U_c^2 + \frac{4 k}{H_0} U_c - \left(\rho g H_0 \frac{\pi}{4} - 2 \tau_0 \right) = 0   \, ,
\label{eq:uc-1}
\end{equation}  
which can be solved for the collapse velocity $U_c$
\begin{equation}
U_c = \frac{-\frac{4k}{H_0} + \sqrt{\left(\frac{4k}{H_0}\right)^2 + \frac{\pi \rho}{2} \left(\rho g H_0\frac{\pi}{4} - 2\tau_0\right)}}{\frac{\pi \rho}{4}}         \, .
\label{eq:uc-2}
\end{equation} 
This equation is used in computing $\mathrm{1/Ga_k}$. By rearranging equation \ref{eq:uc-1}, we find:  
\begin{equation}
1 = \frac{1}{2} \mathrm{Fr} + \frac{16}{\pi} \frac{1}{\mathrm{Ga_k}} + \frac{8}{\pi} \mathrm{Pl}   \, ,
\label{eq:pi-extra-1}
\end{equation}
where the first term on right-hand side is denoted as $\Pi_{\mathrm{Fr}}$, the second term as $\Pi_{\mathrm{Ga_k}}$, and the last term as $\Pi_{\mathrm{Pl}}$. Notably, $\Pi_{\mathrm{Fr}}$, $\Pi_{\mathrm{Ga_k}}$ and $\Pi_{\mathrm{Pl}}$ are bounded between 0 and 1. However, for the flow cases analyzed here, $\Pi_{\mathrm{Fr}} \approx 0$. Equation \ref{eq:pi-extra-1} becomes then
\begin{equation}
1 = \frac{16}{\pi} \frac{1}{\mathrm{Ga_k}} + \frac{8}{\pi} \mathrm{Pl}   \, .
\label{eq:pi-extra-2}
\end{equation}
When $\tau_0 = 0$Pa, $\Pi_{\mathrm{Ga}} = 1$. On the other hand, a fully plastic case occurs at $\Pi_{\mathrm{Pl}} = 1$, indicating ultimately that fluid column behaves as a non-deformable solid (e.g., absence of flow).    

%%%%%%%%%%%%%%%%%%%%%%%%%%%%%%%%%%%%%%%%%%%%%%%%%%%%%%%%%%%%%%%%%%%%%%%%%%%%%%
%%%%%%%%%%%%%%%%%%%%%%%%%%%%%%%%%%%%%%%%%%%%%%%%%%%%%%%%%%%%%%%%%%%%%%%%%%%%%%
\section{Results and Discussion} \label{RD}
\begin{figure*}%[h!]
\centering
\includegraphics[angle=0, scale=0.145]{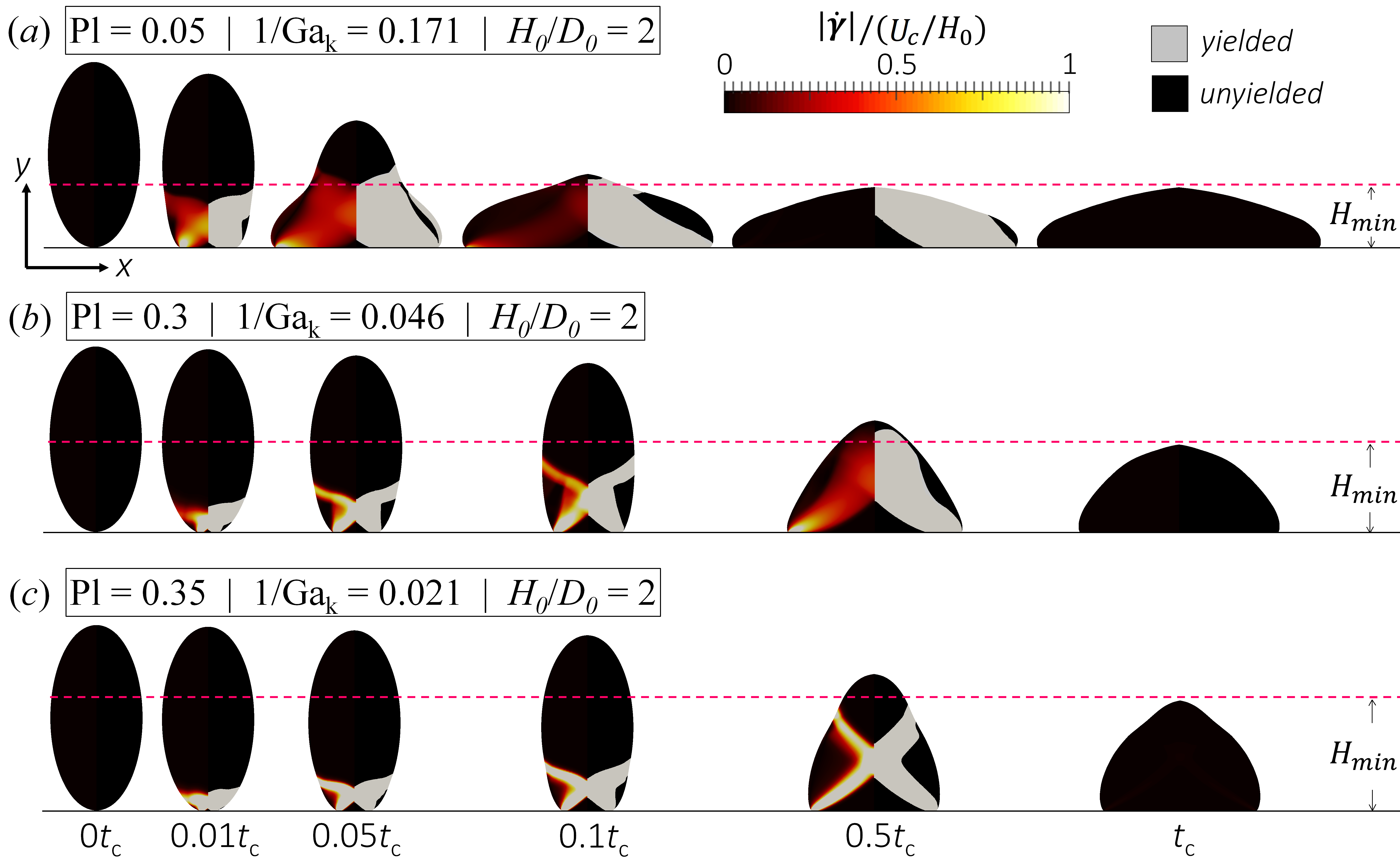}
%\vspace*{-0.5cm}
\caption{Spreading of elliptical objects with $H_0/D_0 =2$, following their collapse over a solid surface for three different $\mathrm{Pl}$-$\mathrm{1/Ga_k}$ pairs: (\textit{a}) $\mathrm{Pl}= 0.05$ , $\mathrm{1/Ga_k} = 0.171$; (\textit{b}) $\mathrm{Pl} = 0.3$, $\mathrm{1/Ga_k} = 0.046$; (\textit{c}) and $\mathrm{Pl} = 0.35$, $\mathrm{1/Ga_k} = 0.021$. Each sub-figure is composed of six images illustrating the spreading of the object at six different times, from onset ($t/t_c =0$) to stoppage ($t/t_c =1$). These images also highlight the contours of the deformation rate $|\boldsymbol{\dot{\gamma}}|$ (made dimensionless by $U_c/H_0$) on the left, together with the unyielded regions ($|\boldsymbol{\tau}| \leq \tau_0$; black areas behaving like an ideal solid) and the yielded regions ($|\boldsymbol{\tau}| > \tau_0$; gray areas behaving like a non-Newtonian liquid) on the right.}
\label{fig-2}
\end{figure*}

%%%%%%%%%%%%%%%%%%%%%%%%%%%%%%%%%%%%%%%%%%%%%%%%%%%%%%%%%%%%%%%%%%%%%%%%%%%%%%
%%%%%%%%%%%%%%%%%%%%%%%%%%%%%%%%%%%%%%%%%%%%%%%%%%%%%%%%%%%%%%%%%%%%%%%%%%%%%%
Figure \ref{fig-2} shows the gravity-induced collapse of elliptical objects whose $H_0/D_0 =2$, following their spreading over a solid surface at three different $\mathrm{Pl}$-$\mathrm{1/Ga_k}$ pairs: $\mathrm{Pl}= 0.05$, $\mathrm{1/Ga_k} = 0.171$ (sub-figure \ref{fig-2}\textit{a}); $\mathrm{Pl} = 0.3$, $\mathrm{1/Ga_k} = 0.046$ (sub-figure \ref{fig-2}\textit{b}); and $\mathrm{Pl} = 0.35$, $\mathrm{1/Ga_k} = 0.021$ (sub-figure \ref{fig-2}\textit{c}). Each sub-figure is composed of six snapshots illustrating the spreading at six different instants, from the onset ($t/t_c =0$) to stoppage ($t/t_c =1$). These images highlight not only the instantaneous morphology of the objects but also the contours of the deformation rate $|\boldsymbol{\dot{\gamma}}|$ (made dimensionless by $U_c/H_0$) on the left, together with the unyielded regions ($|\boldsymbol{\tau}| \leq \tau_0$; black areas behaving like an ideal solid) and the yielded regions ($|\boldsymbol{\tau}| > \tau_0$; gray areas behaving like a non-Newtonian liquid) on the right.

As shown by comparing sub-figures \ref{fig-2}(\textit{a}), \ref{fig-2}(\textit{b}), and \ref{fig-2}(\textit{c}), the increase in plastic number $\mathrm{Pl}$ (conversely a decrease in $\mathrm{1/Ga_k}$, see equation \ref{eq:pi-extra-2}) accentuates yield stress relative to the gravitational stress, resulting in lesser relative spreading $D_{max}/D_0$. Additionally, the increase in $\mathrm{Pl}$ favors the development of unyielded regions (in black), especially in areas of the 2D objects far from the solid surface, where the deformation rate is minimal. It is also worth noting that the most significant values of $ | \boldsymbol{\dot{\gamma}} | /\left(U_c/H_0\right)$ primarily develop within a layer whose thickness is comparable to $H_{min}$, as indicated by the magenta dotted lines. In other words, $H_{min}$ corresponds to the thickness of the layer in the non-Newtonian fluid where the energy dissipation mechanism is predominantly active.

\begin{figure*}%[h!]
\centering
\includegraphics[angle=0, scale=0.31]{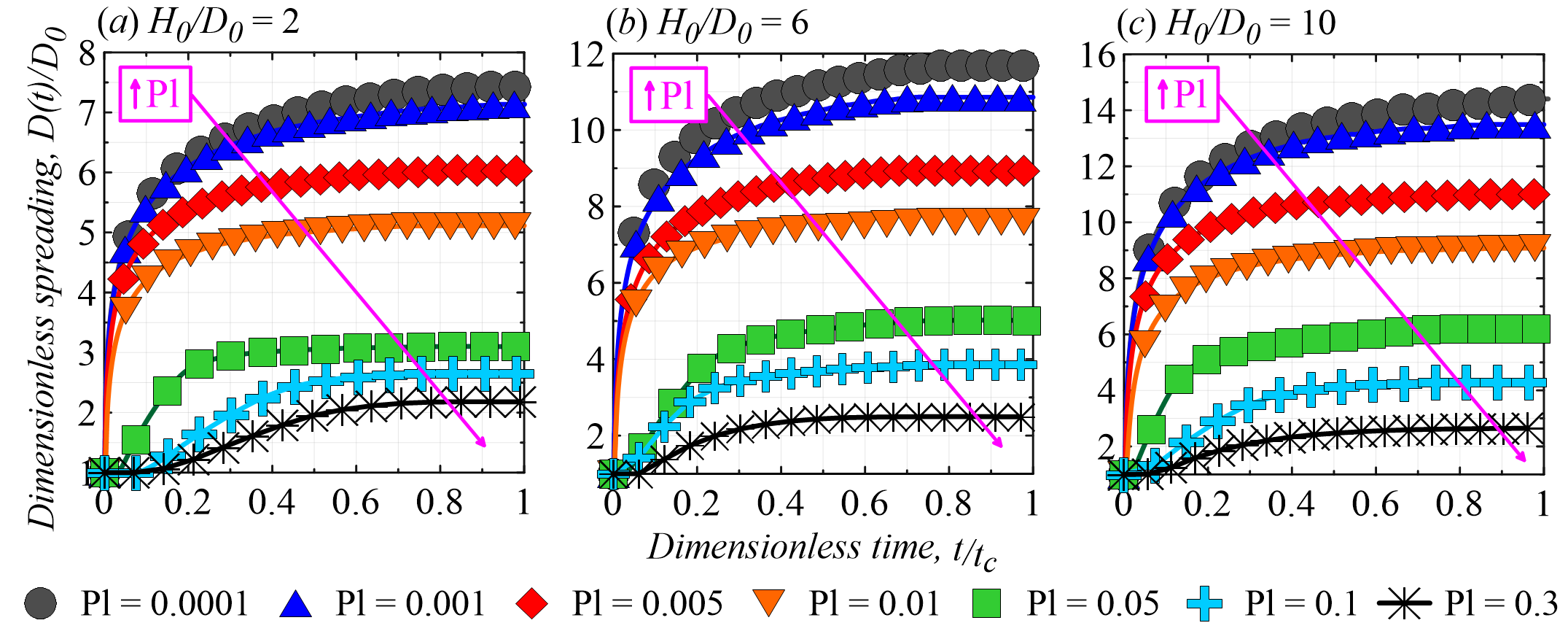}
\vspace*{-0.25cm}
\caption{Instantaneous relative spreading $D(t)/D_0$ as a function of $t/t_c$ to highlight the effect of the aspect ratio: (\textit{a}) $H_0/D_0=2$, (\textit{b}) $H_0⁄D_0=6$, and (\textit{c}) $H_0/D_0=10$. Seven $\mathrm{Pl}$ values are considered in each sub-figure: $\mathrm{Pl} = 0.0001$ (gray circles), $\mathrm{Pl}=0.001$ (blue triangles), $\mathrm{Pl}=0.005$ (red diamonds), $\mathrm{Pl}=0.01$ (orange inverted triangles), $\mathrm{Pl}=0.05$ (green squares), $\mathrm{Pl}=0.1$ (cyan crosses), and $\mathrm{Pl}=0.3$ (black asterisks). Moving downwards on each sub-plot denotes increasing $\mathrm{Pl}$ (indicated by the magenta arrows).}
\label{fig-3}
\end{figure*}

The influence of $H_0/D_0$ on the collapse process is highlighted in figure \ref{fig-3}, where the instantaneous relative spreading $D(t)/D_0$ is plotted as a function of $t/t_c$ at three different aspect ratios:  $H_0/D_0=2$ (sub-figure \ref{fig-3}\textit{a}), $H_0/D_0=6$ (sub-figure \ref{fig-3}\textit{b}), and $H_0/D_0=10$ (sub-figure \ref{fig-3}\textit{a}). Moving downwards on each sub-plot denotes increasing $\mathrm{Pl}$ across seven values represented by distinct symbols (as indicated by the magenta arrows): $\mathrm{Pl}= 0.0001$ (gray circles), $\mathrm{Pl}=0.001$ (blue triangles), $\mathrm{Pl}=0.005$ (red diamonds), $\mathrm{Pl}=0.01$ (orange inverted triangles), $\mathrm{Pl}=0.05$ (green squares), $\mathrm{Pl}=0.1$ (cyan crosses), and $\mathrm{Pl}=0.3$ (black asterisks), which results in lesser relative spreading as previously discussed. As observed by confronting sub-figures \ref{fig-3}(\textit{a}), \ref{fig-3}(\textit{b}), and \ref{fig-3}(\textit{c}), $D(t)/D_0$ increases with $H_0/D_0$, a direct consequence of the augmentation of unyielded regions at the top part of the elliptical objects, ultimately causing deformation to be concentrated in regions closer to the solid surface. Similar trends are seen in elliptical (2D) and prolate (3D) impacting objects \cite{Isukwem-24b,Isukwem-24e, Isukwem-24a}.

\begin{figure*}%[h!]
\centering
\includegraphics[angle=0, scale=0.31]{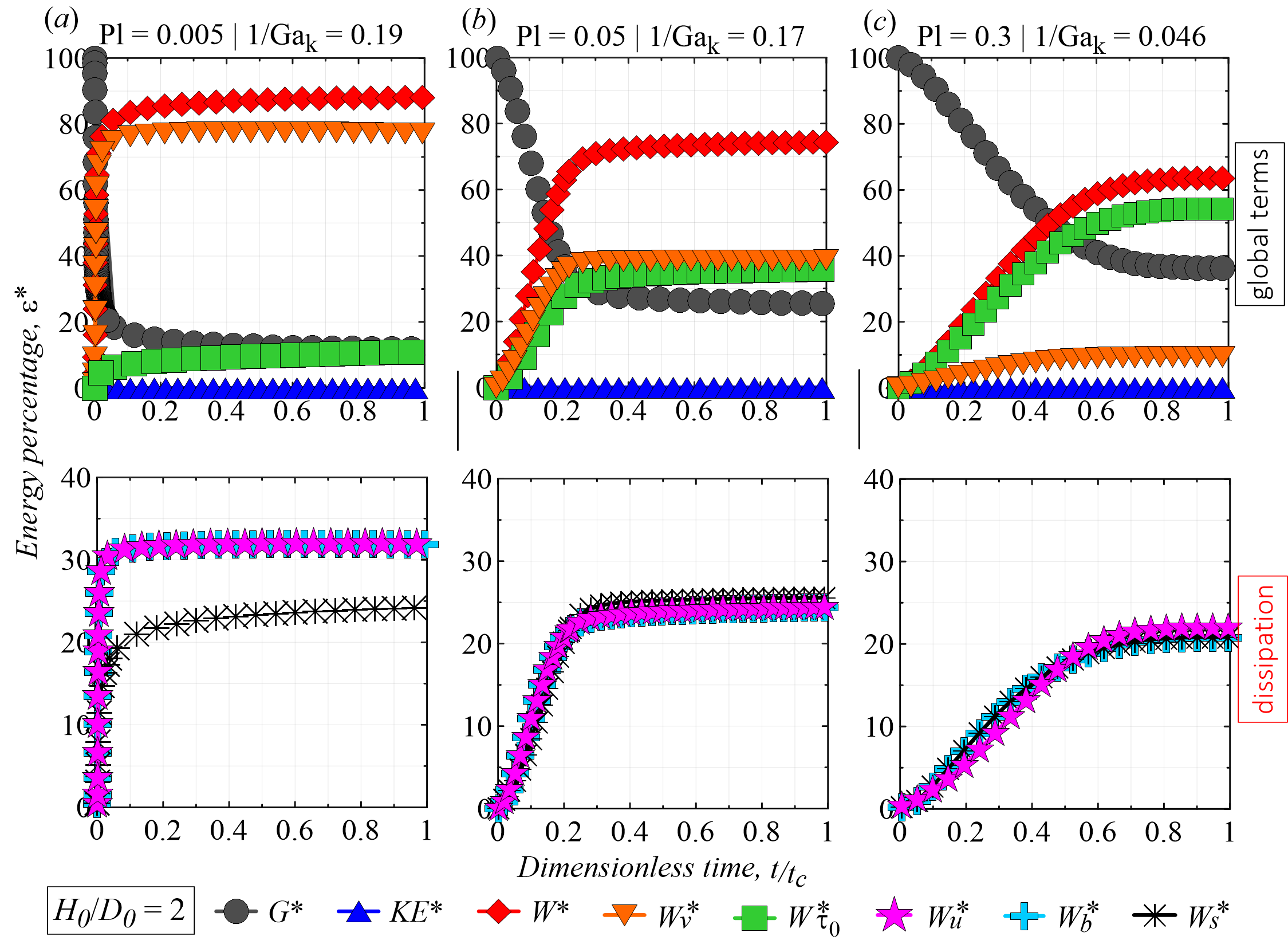}
\vspace*{-0.7cm}
\caption{The energy percentage $\epsilon^{\ast}$ during the collapse and spreading of viscoplastic elliptical objects with $H_0/D_0 = 2$ is presented for three different $\mathrm{Pl}$-$\mathrm{1/Ga_k}$) pairs: (\textit{a}) $\mathrm{Pl} = 0.005$, $\mathrm{1/Ga_k} = 0.19$; (\textit{b}) $\mathrm{Pl}$ = 0.05, $\mathrm{1/Ga_k}$ = 0.17; and (\textit{c}) $\mathrm{Pl}$ = 0.3, $\mathrm{1/Ga_k}$ = 0.046.}
\label{fig-4}
\end{figure*}

To gain deeper insight into the physical mechanisms governing the spreading process, we examine the energy transfer during the collapse of viscoplastic elliptical objects, as illustrated in figure \ref{fig-4}. The energy percentage curves are plotted against dimensionless time, illustrating the gravitational energy $G$, kinetic energy $KE$, and dissipated energy $W$. These energy terms (per unit width) are defined as follows:  
\begin{equation}
G = \int_{S}^{} \rho g y dS    \, ,
\label{eq8.13}
\end{equation}
\begin{equation}
KE = \int_{S}^{} \frac{\rho {|\boldsymbol{u}|}^2}{2} dS    \, ,
\label{eq8.14}
\end{equation}
\begin{equation}
W = \int_{t}^{} \int_{S}^{} \left( \frac{k}{2} {| \boldsymbol{{\dot{\gamma}}} |} + \tau_0 \right) {| \boldsymbol{{\dot{\gamma}}}  |}  dS dt    \, ,
\label{eq8.15}
\end{equation}
where $S$ denotes the surface of the ellipses. The dissipated energy (per unit width) is made up of its viscous contributions $W_v= \int_{t}^{} \int_{S}^{} \frac{k}{2} {| \boldsymbol{{\dot{\gamma}}} |^2}  dS dt$ and yield-stress-based contributions $W_{\tau_0}= \int_{t}^{} \int_{S}^{} \tau_0 {| \boldsymbol{{\dot{\gamma}}}  |}  dS dt$. Moreover, the dissipated energy can be divided into three components: $W_u = \int_{t}^{} \int_{S}^{} \frac{k}{2} \frac{\dot{\gamma}_{yy}^2}{2} + \tau_0 \frac{\dot{\gamma}_{yy}^2}{2 |\boldsymbol{\dot{\gamma}}|} dS dt$ related to the planar compression on the wall-normal ($y$) direction, $W_b = \int_{t}^{} \int_{S}^{} \frac{k}{2} \frac{\dot{\gamma}_{xx}^{2}}{2} + \tau_0 \frac{\dot{\gamma}_{xx}^{2}}{2 |\boldsymbol{\dot{\gamma}}|} dS dt $ associated to the planar expansion in the spreading ($x$) direction, and the complementary shear-based part $W_s = W - (W_u + W_b)$ (where $\dot{\gamma}_{xx} = 2\partial{u_x}/\partial{x}$, and $\dot{\gamma}_{yy} = 2\partial{u_y}/\partial{y}$). Each energy term is normalized by the total system energy $E_0 = G(t=0)$: $G^{\ast} = G/E_0 ~ \times$ 100[\%] (gray circles), $KE^{\ast} = KE/E_0 ~ \times$ 100[\%] (blue triangles), $W^{\ast} = W/E_0 ~ \times$ 100[\%] (red diamonds), $W_{v}^{\ast} = W_{v}/E_0 ~ \times$ 100[\%] (orange inverted triangles), $W_{\tau_0}^{\ast} = W_{\tau_0}/E_0 ~ \times$ 100[\%] (green squares), $W_u^{\ast} = W_u/E_0 ~ \times$ 100[\%] (magenta stars), $W_b^{\ast} = W_b/E_0 ~ \times$ 100[\%] (cyan crosses), and $W_s^{\ast} = W_s/E_0 ~ \times$ 100[\%] (black asterisks). Moreover, each sub-figure corresponds to a specific $\mathrm{Pl}$-$\mathrm{1/Ga_k}$ pair at $H_0/D_0 =$ 2: (\textit{a}) $\mathrm{Pl} =$ 0.005 and $\mathrm{1/Ga_k} =$ 0.19; (\textit{b}) $\mathrm{Pl} =$ 0.05 and $\mathrm{1/Ga_k} =$ 0.17; and (\textit{c}) $\mathrm{Pl} =$ 0.3 and $\mathrm{1/Ga_k} =$ 0.046.

For all cases presented in figure \ref{fig-4} the object's spreading is initiated by its initial gravitational energy, $G^{\ast}(t = 0) =$ 100\%, which is dominantly dissipated by viscoplastic effects, with a negligible part converted into kinetic energy (as anticipated in section \ref{PFNMDN}). At $\mathrm{Pl} =$ 0.005 and $\mathrm{1/Ga_k} =$ 0.19 (sub-figure \ref{fig-4}\textit{a}), $G^{\ast}$ is mainly dissipated by viscous effects $W_v^{\ast}$, indicating that the collapse and subsequent spreading is governed by the interplay between gravitational and viscous stresses, characteristic of a gravito-viscous regime. As $\mathrm{Pl}$ increases (and consequently $\mathrm{1/Ga_k}$ decreases), the yield-stress contribution to the dissipation $W_{\tau_0}$ becomes significant (sub-figure \ref{fig-4}\textit{b}), eventually becoming the dominant dissipation mechanism in the spreading (sub-figure \ref{fig-4}\textit{c}). In this gravito-plastic regime, the object's deformation is driven by the competition between gravitational and yield-stress effects.

Focusing on the dissipation terms displayed in figure \ref{fig-4}, it is essential to note that $W_u^{\ast}$, $W_b^{\ast}$ and $W_s^{\ast}$ are comparable in magnitude in all flow regimes (although in the gravito-viscous scenario, the shear component is slightly lower). Expressed differently, the 2D dam breaks explored here undergo a complex dissipative process emerging from equivalent planar compression/expansions-based and shear-based contributions. 

The energy balances depicted in figure \ref{fig-4} reveal the presence of at least two distinct collapse regimes: the gravito-viscous regime (sub-figure \ref{fig-4}\textit{a}); and the gravito-plastic regime (figure \ref{fig-4}\textit{c}). In the gravito-viscous regime, the gravitational energy (per unit width) of 2D objects at collapse [$\sim \rho g H_0 \left( H_0 D_0 \right)$] is primarily dissipated by viscous effects. Since $W_u^{\ast} \approx W_b^{\ast} \approx W_s^{\ast}$ as shown in figure \ref{fig-4}, we choose to define the dissipated energy in the gravito-viscous regime based on a shear perspective [$\sim k\left(\frac{U_c}{H_{min}} \right)D_{max}^2$, where $H_{min}$ is taken as the dissipative length scale, following the deformation rate contours pointed out by figure \ref{fig-2}]. In the gravito-plastic regime, the gravitational energy (per unit width) is dissipated by yield-stress effects ($\sim \tau_0 D_{max}^2$). By equating these energy terms and considering that $H_{min} \sim H_0 D_0/D_{max}$ based on the principle of mass conservation, we can derive the following results:  
\begin{equation}
\frac{D_{max}}{D_0} \sim \left( \mathrm{Ga_k} \frac{H_0}{D_0}\right)^{1/3} ~~~ \text{(gravito-viscous collapse)}\, ,
\label{eq:gravito-viscous_D}
\end{equation}
\begin{equation}
\frac{D_{max}}{D_0} \sim \left( \frac{1}{\mathrm{Pl}} \frac{H_0}{D_0}  \right)^{1/2} ~~~ \text{(gravito-plastic collapse)}\, .
\label{eq:gravito-plastic_D}
\end{equation}
The transition between these two regimes can be determined by equating equations \ref{eq:gravito-viscous_D} and \ref{eq:gravito-plastic_D}, leading to the expression $\left(  \frac{1}{\mathrm{Pl}}  \right)^{1/2}  \left(  \frac{1}{\mathrm{Ga_k}}  \right)^{1/3}  \left(  \frac{H_0}{D_0}  \right)^{1/6} \sim 1$. This dimensionless quantity is referred to as the \textit{collapse number}, denoted as $\mathrm{Co} = \left(  \frac{1}{\mathrm{Pl}}  \right)^{1/2}  \left(  \frac{1}{\mathrm{Ga_k}}  \right)^{1/3}  \left(  \frac{H_0}{D_0}  \right)^{1/6}$. When $\mathrm{Co} \sim 1$, the transition between the gravito-viscous and gravito-plastic regimes is expected to occur.

\begin{figure*}%[h!]
\centering
\hspace*{-1.5cm}
\includegraphics[angle=0, scale=0.41]{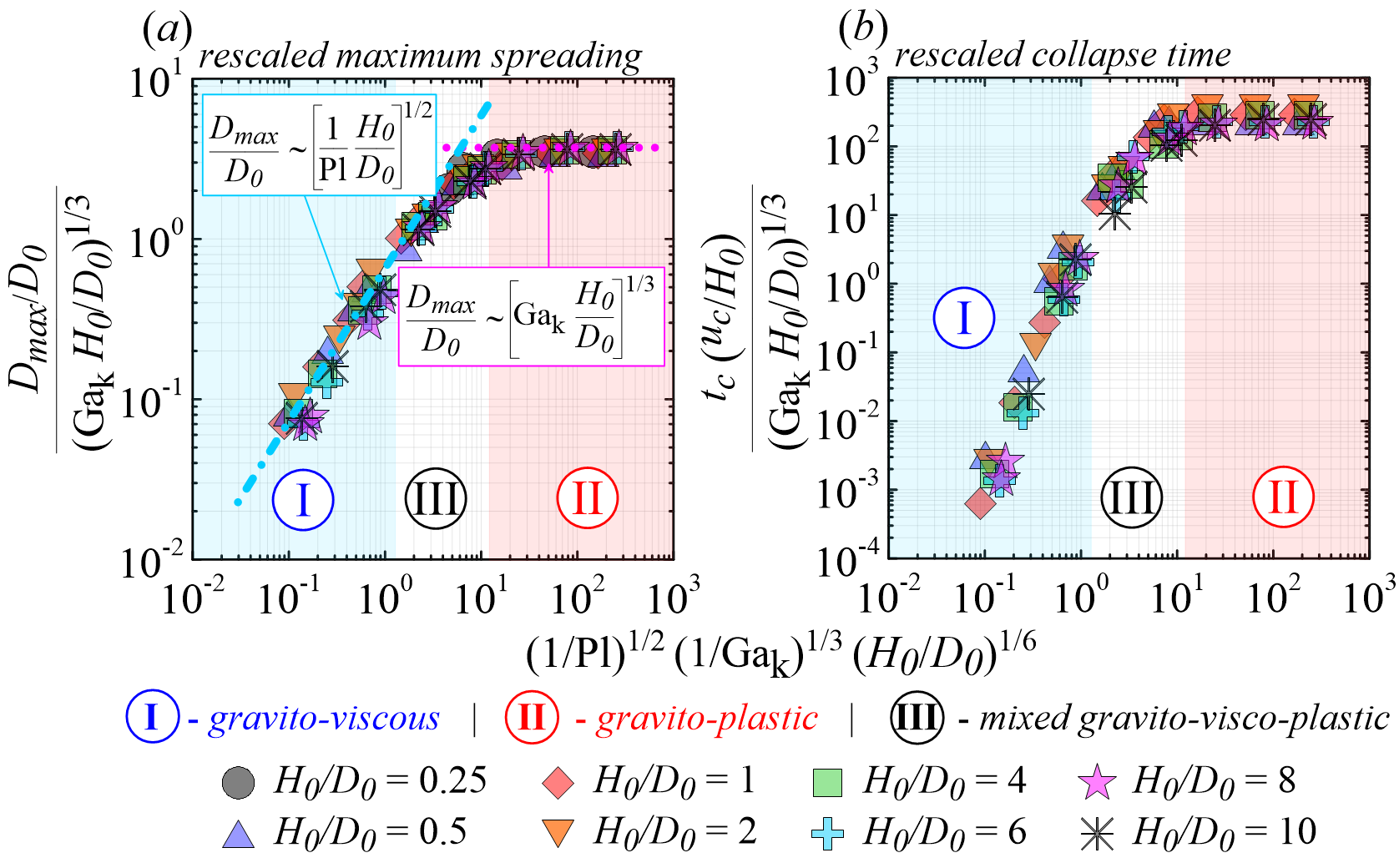}
\vspace*{-0.25cm}
\caption{(\textit{a}) Rescaled maximum spreading $ \frac{ D_{max}/D_0}{\left( \mathrm{Ga_k} H_0/D_0 \right)^{1/3}} $ plotted against the collapse number $\left(  \frac{1}{\mathrm{Pl}}  \right)^{1/2}  \left(  \frac{1}{\mathrm{Ga_k}}  \right)^{1/3}  \left(  \frac{H_0}{D_0}  \right)^{1/6}$. (\textit{b}) Rescaled collapsing time $ \frac{ t_c(U_c/H_0) }{\left( \mathrm{Ga_k} H_0/D_0 \right)^{1/3}} $ against the collapse number. The numerical data align along a master curve that is divided three distinct regions: (I) the blue zone, which corresponds to the gravito-plastic regime; (II) the red zone, which is associated to the gravito-viscous regime; and (III) the white zone, indicating a mixed gravito-visco-plastic regime. Each symbol is associated with an aspect ratio.}
\label{fig-5}
\end{figure*}

The theoretical argument stressed above is supported by the data shown in sub-figure \ref{fig-5}(\textit{a}), where the rescaled maximum spreading $ \frac{ D_{max}/D_0}{\left( \mathrm{Ga_k} H_0/D_0 \right)^{1/3}} $, is plotted against the collapse number $\left(  \frac{1}{\mathrm{Pl}}  \right)^{1/2}  \left(  \frac{1}{\mathrm{Ga_k}}  \right)^{1/3}  \left(  \frac{H_0}{D_0}  \right)^{1/6}$. The results reveal a consistent trend line that can be categorised into three distinct regions: (I) the blue region, which corresponds to the gravito-plastic regime where $\frac{D_{max}}{D_0} \sim \left(\frac{1}{\mathrm{Pl}} \frac{H_0}{D_0}  \right)^{1/2}$ (blue dash-dotted line); (II) the red region, associated with the gravito-viscous regime, where $\frac{D_{max}}{D_0} \sim \left( \mathrm{Ga_k} \frac{H_0}{D_0}\right)^{1/3}$ (magenta dotted line); and (III) the white region, representing a mixed gravito-visco-plastic regime for which the gravitational, viscous and yield stresses all play a significant role (in this case, $W_v^* \approx W_{\tau_0}^*$ as indicated in sub-figure \ref{fig-4}\textit{b}; $1 \lesssim \mathrm{Co} \lesssim 10$).

The rescaled collapse time $ \frac{ t_c(U_c/H_0) }{\left( \mathrm{Ga_k} H_0/D_0 \right)^{1/3}} $ is plotted in sub-figure \ref{fig-5}(\textit{b}) as a function of $\mathrm{Co}$. The collapse time is approximated as $t_c \sim (H_0 - H_{min})/U_c$, which allows us to write that $t_c(U_c/H_0) \sim (1 - H_{min}/H_0)$. As a result, we find that $t_c(U_c/H_0)$ is a function of $H_{min}/H_0$, which in turn is a function $\mathrm{Co}$ since $H_{min}/H_0 \sim D_0/D_{max}$ by mass conservation. Consequently, $t_c(U_c/H_0)$ is a function of $\mathrm{Co}$, as pointed out by sub-figure \ref{fig-5}(\textit{b}) where the numerical data collapse across a master curve.

Lastly, it is essential to emphasize that the new scaling laws presented here are different from those previously reported in the literature using cylindrical and prismatic columns \citep[for which $D_{max}/D_0$ scales with ${\left( \mathrm{Ga_k} \frac{H_0}{D_0} \right)^{1/5}}$ in the gravito-viscous regime and ${\left(  \frac{1}{\mathrm{Pl}}  \frac{H_0}{D_0} \right)^{1/3}}$ in the gravito-plastic regime;][]{Valette-21}. Rationally, these differences emerge from the two-dimensional nature of our flow cases, which ultimately affects the dissipative mechanism within the collapsing objects. In other words, axisymmetric-based scaling laws are not able to predict both the maximum spreading and the collapsing time related to the dam break of the 2D viscoplastic columns. These differences, therefore, allow us to assume the existence of a change in scaling laws induced by the transition from 3D-like to 2D-like objects. However, this transition remains to be studied in depth in future works.

%%%%%%%%%%%%%%%%%%%%%%%%%%%%%%%%%%%%%%%%%%%%%%%%%%%%%%%%%%%%%%%%%%%%%%%%%%%%%%
%%%%%%%%%%%%%%%%%%%%%%%%%%%%%%%%%%%%%%%%%%%%%%%%%%%%%%%%%%%%%%%%%%%%%%%%%%%%%%
\section{Concluding Remarks}
%%%%%%%%%%%%%%%%%%%%%%%%%%%%%%%%%%%%%%%%%%%%%%%%%%%%%%%%%%%%%%%%%%%%%%%%%%%%%%
%%%%%%%%%%%%%%%%%%%%%%%%%%%%%%%%%%%%%%%%%%%%%%%%%%%%%%%%%%%%%%%%%%%%%%%%%%%%%%
%\vspace*{-0.5cm}
In this short communication, we have presented a theoretical and numerical study devoted to the physical mechanisms driving the dam break of viscoplastic elliptical columns on a sticky solid (absence of sliding) under gravity effects. Our 2D numerical simulations were based on a variational multi-scale approach dedicated to non-Newtonian multiphase flows. The results obtained were analyzed considering the spreading dynamics, energy balances and new scaling laws. 

Three collapse regimes have been highlighted: the gravito-viscous regime, resulting from a competition between gravitational and viscous stresses, for which $D_{max}/D_0$ scales with  ${\left( \mathrm{Ga_k} \frac{H_0}{D_0} \right)^{1/3}}$; the gravito-plastic regime, resulting from a balance between gravitational and yield stresses, for which $D_{max}/D_0$ scales with ${\left(  \frac{1}{\mathrm{Pl}}  \frac{H_0}{D_0} \right)^{1/2}}$; and the mixed inertio-visco-plastic regime, for which inertial, viscous and yield stresses are all important. These results were synthesized in the form of a master curve giving the maximum spreading of viscoplastic elliptical objects as a function of a unique dimensionless number called collapse number $\mathrm{Co} = \left(  \frac{1}{\mathrm{Pl}}  \right)^{1/2}  \left(  \frac{1}{\mathrm{Ga_k}}  \right)^{1/3}  \left(  \frac{H_0}{D_0}  \right)^{1/6}$ relating the three important dimensionless parameters of the problem, namely $\mathrm{Ga_k}$, $\mathrm{Pl}$ and $H_0/D_0$. The collapse time appears as a function of $\mathrm{Co}$. Interestingly, regardless of the collapsing regime, the explored 2D dam breaks undergo a complex dissipative process emerging from equivalent planar compression/expansion-based and shear-based contributions. 

The scaling laws presented here for elliptical objects being different from those associated with the collapse of cylindrical and prismatic viscoplastic columns \citep{Valette-21}, it would be interesting to analyze in future works the physical mechanisms dominating the spreading transition of objects from 3D to 2D (3D-to-2D transition). Finally, it would also be interesting to incorporate other sources of complexities, including thixotropy, elasticity, inertia, surface tension and slip laws.       

%%%%%%%%%%%%%%%%%%%%%%%%%%%%%%%%%%%%%%%%%%%%%%%%%%%%%%%%%%%%%%%%%%%%%%%%%%%%%%
%%%%%%%%%%%%%%%%%%%%%%%%%%%%%%%%%%%%%%%%%%%%%%%%%%%%%%%%%%%%%%%%%%%%%%%%%%%%%%
\textit{\textbf{Acknowledgements:}} 
%%%%%%%%%%%%%%%%%%%%%%%%%%%%%%%%%%%%%%%%%%%%%%%%%%%%%%%%%%%%%%%%%%%%%%%%%%%%%%
%%%%%%%%%%%%%%%%%%%%%%%%%%%%%%%%%%%%%%%%%%%%%%%%%%%%%%%%%%%%%%%%%%%%%%%%%%%%%%
The authors would like to acknowledge the support from the UCA$^\mathrm{JEDI}$ program (IDEX of Universit\'e C\^ote d'Azur), the PSL Research University under the program `Investissements d'Avenir' launched by the French Government and implemented by the French National Research Agency (ANR) with the reference ANR-10-IDEX-0001-02 PSL, and the ANR for supporting the INNpact project under the `Jeunes chercheuses et jeunes chercheurs' program.   

%\newpage
%\clearpage

%%%%%%%%%%%%%%%%%%%%%%%%%%%%%%%%%%%%%%%%%%%%%%%%%%%%%%%%%%%%%%%%%%%%%%%%%%%%%%
%%%%%%%%%%%%%%%%%%%%%%%%%%%%%%%%%%%%%%%%%%%%%%%%%%%%%%%%%%%%%%%%%%%%%%%%%%%%%%
%\bibliographystyle{./apsrev4-2}
\bibliography{dambreaks-ref}% Produces the bibliography via BibTeX.
%%%%%%%%%%%%%%%%%%%%%%%%%%%%%%%%%%%%%%%%%%%%%%%%%%%%%%%%%%%%%%%%%%%%%%%%%%%%%%
%%%%%%%%%%%%%%%%%%%%%%%%%%%%%%%%%%%%%%%%%%%%%%%%%%%%%%%%%%%%%%%%%%%%%%%%%%%%%%
\end{document}